\documentclass[%
 reprint,
 amsmath,amssymb,
 aps,
prb,
]{revtex4-2}

\usepackage{graphicx}
\usepackage{dcolumn}
\usepackage{bm}

\usepackage[utf8]{inputenc}
\usepackage[T1]{fontenc}
\usepackage{booktabs, array, mathptmx, tabularx, booktabs, lipsum, amsmath,multirow}
\usepackage{siunitx, xcolor}
\usepackage{newunicodechar}
\newunicodechar{δ}{\delta}
\usepackage[version=4]{mhchem}

\usepackage{appendix}
\usepackage{subcaption}
\usepackage[percent]{overpic}
\usepackage{adjustbox}
\usepackage{epstopdf}
\usepackage{xcolor}
\usepackage{braket}
\usepackage[colorlinks,linkcolor=blue,anchorcolor=blue,citecolor=blue]{hyperref}

\begin{document}
\title{Many-body mobility edges in one dimension revealed by efficient and interpretable feature-based learning with Kolmogorov-Arnold Networks}
\author{Siqi Dai}
\author{Tian-Cheng Yi}
\author{Xingbo Wei}
\author{Yunbo Zhang}
\email{Contact author: ybzhang@zstu.edu.cn}
\affiliation{Zhejiang Key Laboratory of Quantum State Control and Optical Field Manipulation, Department of Physics, Zhejiang Sci-Tech University, 310018 Hangzhou, China}
\begin{abstract}
We study the many-body localization (MBL) transition in interacting fermionic systems on disordered one-dimensional lattices using a physics-informed machine-learning framework. Instead of feeding full many-body wave functions into the model, we construct a compact feature representation based on four physically motivated observables: the inverse participation ratio, the Shannon entropy, the many-body hybridization parameter, and the mean level-spacing ratio. These quantities capture complementary aspects of localization, entanglement, and spectral correlations, and are used to train a Kolmogorov--Arnold Network (KAN) classifier on eigenstates deep in the weak and strong disorder regimes.
The resulting KAN achieves a validation accuracy exceeding $99.9\%$, comparable to that of convolutional neural networks trained directly on high-dimensional wave-function data, while requiring substantially reduced input dimensionality and significantly shorter training time. Applying the trained classifier across the full energy spectrum yields energy-resolved phase diagrams that reveal a clear many-body mobility edge and provide a consistent estimate of the critical disorder strength. The approach is inherently extensible: additional physically relevant observables can be incorporated into the feature space in a systematic manner without altering the overall architecture. Our results demonstrate that feature-based learning with KAN provides an efficient, scalable, and interpretable methodology for identifying many-body localization transitions, offering a practical alternative to raw-data-based neural network approaches.
\end{abstract}

\maketitle

\section{\label{sec:introduction}Introduction}
In recent years, neural network models have demonstrated considerable potential in condensed matter physics research. Such models are capable of automatically extracting universal features from large-scale datasets and effectively generalizing to unseen data scenarios. They have been widely employed in tasks such as quantum state reconstruction, acceleration of first-principles calculations, and classification of phases of matter based on numerical simulations and experimental data \cite{Carrasquilla2017,PhysRevB.94.195105,PhysRevE.96.022140,PhysRevLett.98.146401,Torlai2018,doi:10.1126/science.aag2302}. Owing to their powerful function approximation capabilities, neural networks can also serve as variational representations of many-body quantum states, achieving accuracy in ground-state energy calculations comparable to that of leading conventional methods \cite{Carrasquilla2017,doi:10.1126/science.aag2302}.

This capability positions neural networks as a powerful tool for investigating fundamental problems in quantum many-body physics, such as thermalization and localization. According to the eigenstate thermalization hypothesis (ETH)  \cite{PhysRevA.43.2046,PhysRevE.50.888,Mark_Srednicki_1996}, a system in the ergodic phase thermalizes, causing local observables to eventually match thermal expectation values \cite{D'Alessio03052016,Deutsch_2018,nature452_854}. In contrast, under strong disorder, a system may enter a many-body localization (MBL) phase. In this regime, thermalization is suppressed and information about the initial state can persist for long times \cite{BASKO20061126,PhysRevB.76.052203,PhysRevB.75.155111,PhysRevB.77.064426,ALET2018498,PhysRevLett.95.206603,PhysRevB.82.174411,doi:10.1126/science.aaa7432}. This phenomenon has been observed experimentally in ultracold atomic systems and suggests potential applications in quantum information storage \cite{doi:10.1126/science.aaf8834,annurev:/content/journals/10.1146/annurev-conmatphys-031214-014726,RevModPhys.91.021001}.
\begin{figure*}[tp]
  \centering
  \begin{subfigure}[b]{0.32\textwidth}
    \centering
    \begin{overpic}[width=\textwidth]{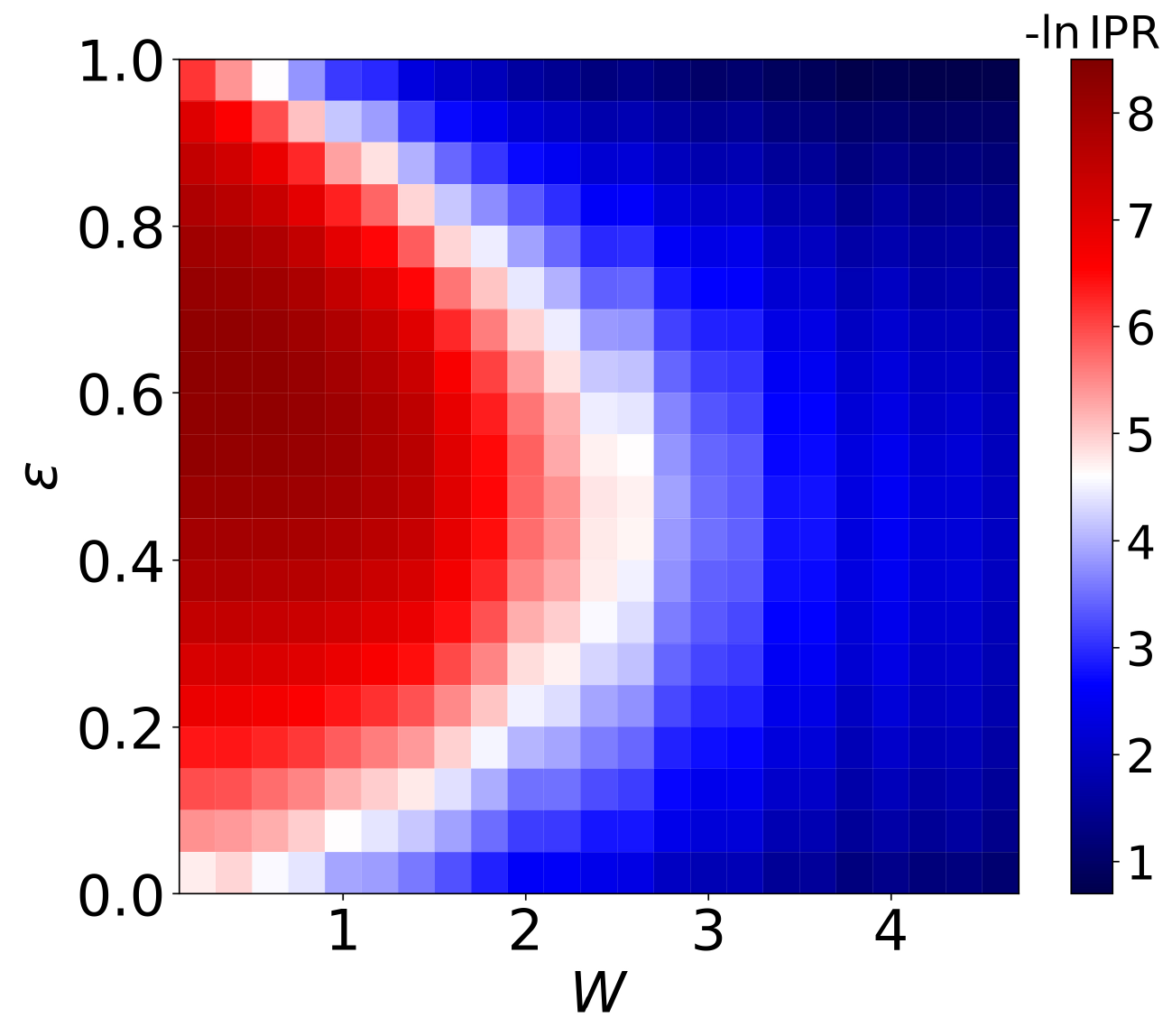}
      \put(77,77){\fontsize{12}{14}\selectfont\color{white}(a)}
    \end{overpic}
    \label{fig:phase-ipr}
  \end{subfigure}
  \begin{subfigure}[b]{0.335\textwidth}
    \centering
    \begin{overpic}[width=\textwidth]{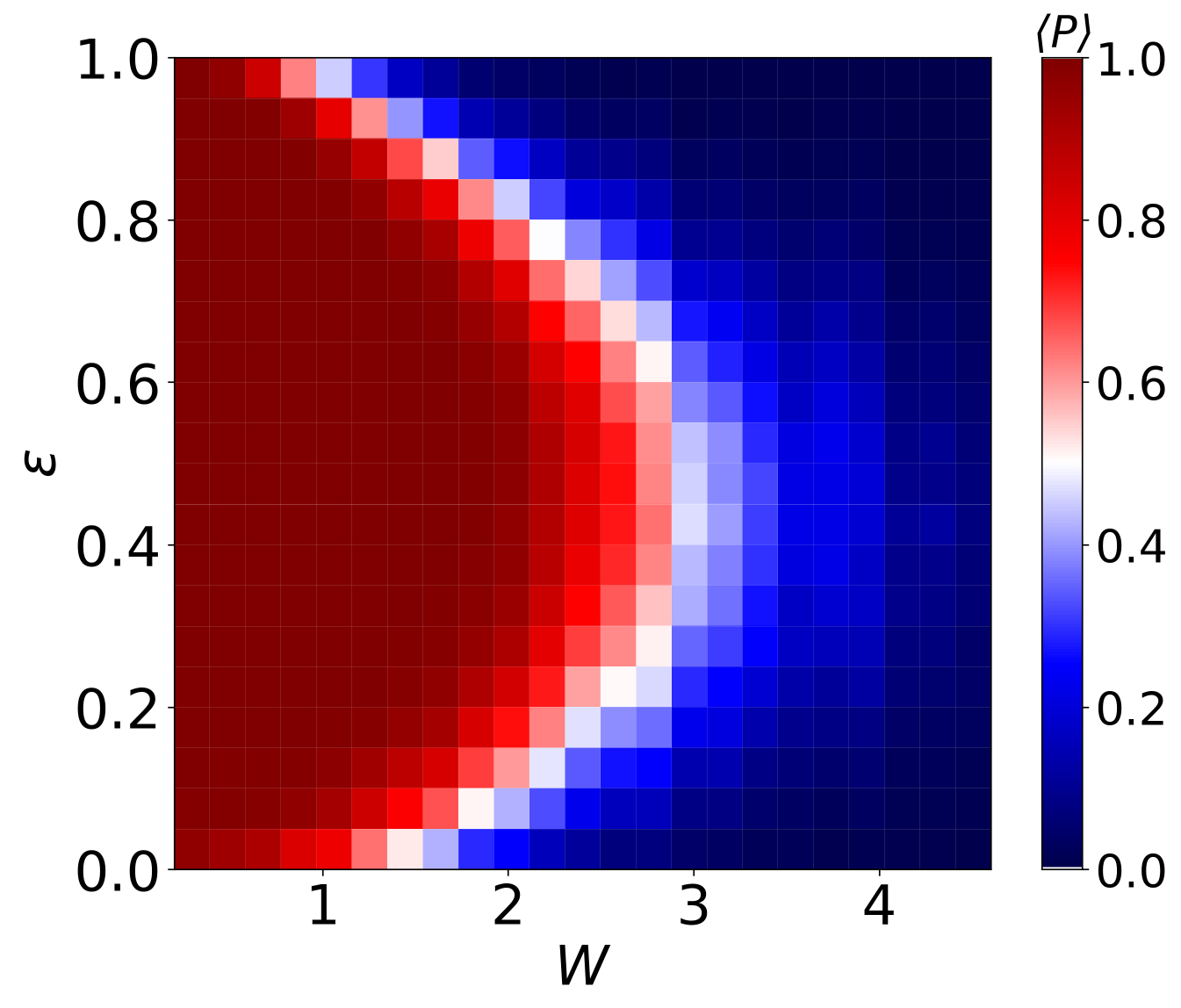}
      \put(75,75){\fontsize{12}{14}\selectfont\color{white}(b)}
    \end{overpic}
    \label{fig:phase-kan}
  \end{subfigure}
  \begin{subfigure}[b]{0.335\textwidth}
    \centering
    \begin{overpic}[width=\textwidth]{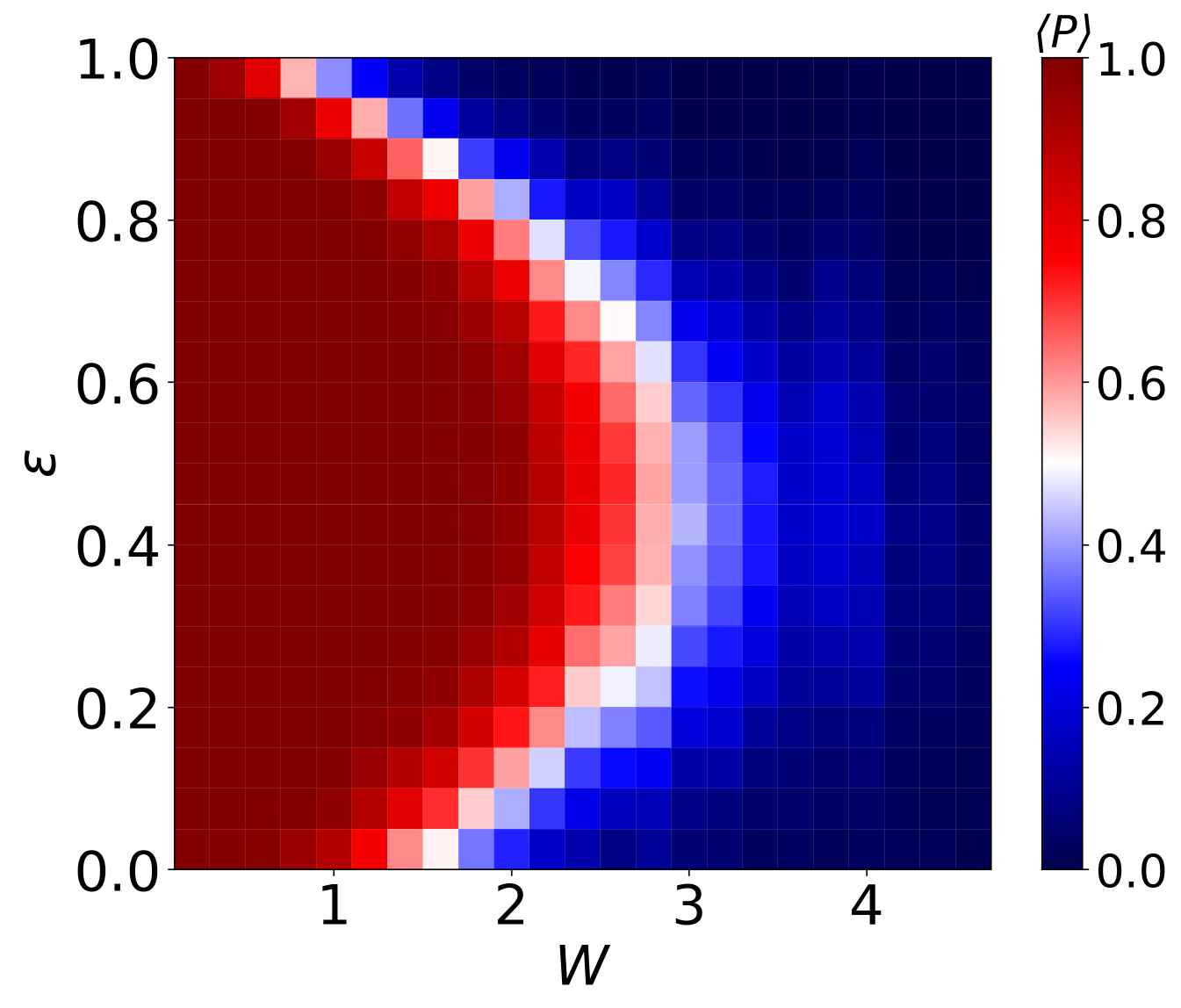}
        \put(75,75){\fontsize{12}{14}\selectfont\color{white}(c)}
    \end{overpic}
    \label{fig:phase-cnn}
  \end{subfigure}
  \caption{Comparison of phase diagrams obtained using different approaches. (a) Phase diagram based on the pure numerical results of inverse participation ratio (IPR). (b) Phase diagram predicted by the Kolmogorov--Arnold Network (KAN) classifier trained on four physically motivated features. (c) Phase diagram obtained using the convolutional neural network (CNN) trained on probability densities in the occupation-number basis states.}
  \label{fig:phase-comparison}
  
\end{figure*}
As a frontier topic in nonequilibrium quantum physics, the study of thermalization and MBL has therefore attracted extensive attention. A conventional approach to investigate the ergodic-MBL transition is to perform exact diagonalization to obtain the eigenvalues and eigenstates of the full spectrum. Several derived diagnostic observables, such as level statistics \cite{PhysRevB.75.155111,PhysRevLett.110.084101}, the inverse participation ratio \cite{Wegner1980,PhysRevB.83.184206}, and entanglement entropy \cite{6773024,LESNE_2014}, serve as indicators to distinguish thermal and MBL phases. The identification of the transition point typically relies on finite-size scaling analysis \cite{PhysRevB.91.081103,PhysRevE.102.062144}. In this procedure, observables are evaluated for different system sizes and then analyzed using appropriate scaling forms. 
However, different scaling assumptions may lead to qualitatively different conclusions. For instance, within the framework of a power-law divergence of the correlation length, the critical disorder strength is estimated to be approximately $W_c\sim 3.7$ \cite{PhysRevB.91.081103}. In contrast, within the framework of the Berezinskii–Kosterlitz–Thouless transition (BKT) scaling theory, the transition point from the thermal phase to the MBL phase drifts toward larger values as the system size increases \cite{PhysRevE.102.062144}. In this context, machine learning techniques have recently been introduced into the study of MBL, opening a new research direction. By extracting hidden structures from complex datasets that are difficult to analyze using conventional approaches, machine learning provides powerful data-analysis tools for characterizing the MBL phase and its transition. Recent work has demonstrated that machine-learning classifiers can be trained to automatically identify distinct phases using data generated from exact diagonalization of model Hamiltonians \cite{Carrasquilla2017,vanNieuwenburg2017,PhysRevB.94.195105,doi:10.7566/JPSJ.85.123706,2v76-dmg6}. Various types of physical data have been used as inputs, including many-body energy spectra \cite{PhysRevLett.121.245701}, wave functions \cite{Carrasquilla2017}, and entanglement spectra \cite{PhysRevB.95.245134} obtained from one-dimensional disordered quantum models. Both supervised and unsupervised machine-learning methods, such as support vector machines and neural networks, have been applied to identify and characterize the ergodic-MBL transition \cite{PhysRevB.94.195105,PhysRevE.96.022140,vanNieuwenburg2017}. An important advantage of these methods is that they can identify the critical point without explicit finite-size scaling analysis, thereby offering an efficient way to map out phase diagrams.

In this study, we employ a feature-based machine learning framework to investigate the ergodic-MBL transition in interacting fermions on a one-dimensional disordered lattice with 16 sites. We move away from conventional approaches that directly use wave functions as model inputs. Instead, we extract a set of physically motivated quantities from each eigenstate, including the inverse participation ratio ($\ln{\mathrm{IPR}}$) \cite{Wegner1980,PhysRevB.83.184206}, the Shannon entropy ($S$) \cite{6773024,LESNE_2014}, the many-body hybridization parameter $g$ \cite{PhysRevX.5.041047,PhysRevLett.123.090603}, and the mean level spacing ratio ($\langle r\rangle$) \cite{PhysRevB.75.155111,PhysRevLett.110.084101}. These features are used as inputs to a Kolmogorov–Arnold Network (KAN) \cite{liu2025kan}, which provides a structured and interpretable mapping for input data.

Our work is intended as a comparison between two methodological paradigms rather than a direct competition between network architectures. Specifically, we contrast a physics-informed approach that incorporates domain knowledge through feature extraction and employs KAN for classification, with a data-driven approach in which convolutional neural network (CNN) \cite{PhysRevB.109.075124,wjvx-5nk7,PhysRevResearch.7.013094,Broecker2017,PhysRevB.99.121104,Cong2019,PhysRevResearch.4.013231} learn representations directly from high-dimensional wave functions. This comparison allows us to assess the extent to which physically meaningful features capture the essential information required for phase classification, without explicit processing of the full wave function.

To implement this comparative study, we proceed as follows. We first train the KAN classifier using eigenstates deep within the ergodic and MBL regimes, and subsequently apply the trained model to intermediate disorder strengths. By averaging predictions over multiple disorder realizations, we construct an energy-resolved phase diagram that reveals the many-body mobility edge, which separates the ergodic and MBL phases in the spectrum. The resulting phase diagram is systematically compared with those obtained from the conventional inverse participation ratio and from the CNN-based approach, as shown in Fig.~\ref{fig:phase-comparison}. Details of the KAN model and training procedure are presented in Sec.~\ref{Machine Learning Frameworks}, while the implementation of the CNN is described in Appendix~\ref{cnn_appendix}.

The remainder of the paper is structured as follows. In Sec.~\ref{sec:models}, we introduce the one-dimensional disordered fermionic model considered in this work. Section~\ref{kan} presents the architecture and underlying principles of the Kolmogorov–Arnold Network (KAN) used in the feature-based classification. The generation of training data via exact diagonalization and the corresponding data preprocessing are described in Sec.~\ref{input_data}, while the training procedure is outlined in Sec.~\ref{training}. Our main results, including the machine-learned phase diagrams, are reported in Sec.~\ref{diagram}. We conclude in Sec.~\ref{conclusion} with a summary of the results and a discussion of their physical implications.

\section{\label{sec:models}Model}

We consider a one-dimensional (1D) lattice system of interacting spinless fermions subject to a disordered external potential. The particles reside on a lattice and exhibit mutual repulsion. The system is described by the many-body Hamiltonian \cite{PhysRevB.26.5033}:
\begin{equation}\label{eq1}
\begin{split}
H = & \sum_{\langle i,j\rangle} \left[ -t \left( c_i^\dagger c_j + c_j^\dagger c_i \right) + V \left(n_i - \frac{1}{2} \right) \left(n_j - \frac{1}{2} \right) \right] \\
& + \sum_{i=1}^N u_i \left(n_i - \frac{1}{2} \right),
\end{split}
\end{equation}
where $c_i^\dagger$ and $c_i$ are the creation and annihilation operators for a fermion at site $i$, respectively, and $n_i = c_i^\dagger c_i$ represents the particle number operator. The notation $\langle i,j \rangle$ indicates a sum over all pairs of adjacent sites. The parameter $t$ denotes the hopping amplitude, while $V$ characterizes the strength of the nearest-neighbor repulsive interaction. In this model, the on-site disorder is incorporated by introducing random potentials $u_i$ at each site. These potentials are independent random variables, each drawn from a uniform distribution within the interval $[-W, W]$, where $W$ represents the strength of the disorder. The shift $n_i \rightarrow n_i - \frac{1}{2}$ ensures particle--hole symmetry at half filling and removes trivial energy offsets, so that the disorder term corresponds to a random field with zero mean in the spin representation.

At a specific point in the parameter space, namely $t = \frac{1}{2}$ and $V = 1$, the fermionic model defined in Eq.~(\ref{eq1}) can be exactly mapped onto a well-known spin model. By applying the Jordan-Wigner transformation \cite{Jordan1928}, i. e. define $c_i=\exp\left(i\pi \prod_{k<i} n_k\right) S_i^-$  with $n_i=S_i^z + \frac{1}{2}$ and $S_i^\pm =S_i^x \pm iS_i^y$, the Hamiltonian (\ref{eq1}) becomes equivalent to the 1D spin-$\frac{1}{2}$ antiferromagnetic Heisenberg chain, which is subjected to a random magnetic field along the $z$-axis \cite{Heisenberg1928}
\begin{equation}\label{eq2}
H = \sum_{\langle i,j\rangle} \mathbf{S}_i \cdot \mathbf{S}_j + \sum_{i=1}^N u_i S_i^z,
\end{equation}
where $\mathbf{S}_i=(S_i^x,S_i^y,S_i^z)$ denotes the vector of spin-$\frac{1}{2}$ operators at site $i$. The detailed derivation of the mapping from the fermionic Hamiltonian in Eq.~(\ref{eq1}) to the spin model in Eq.~(\ref{eq2}) based on the Jordan--Wigner transformation is presented in Appendix~\ref{app:JW}. This spin model is a canonical example to study the many-body localization (MBL) transition as the disorder strength is increased  \cite{PhysRevB.82.174411,PhysRevE.102.062144, RevModPhys.91.021001,PhysRevB.109.075124,PhysRevB.94.045111,PhysRevB.96.104201,PhysRevLett.119.075702}. It is well-established that this system undergoes a transition from an ergodic to an MBL phase. At half filling, corresponding to a total magnetization $\sum_i S_i^{z} = 0$, the critical strength of disorder for this transition is initially estimated to be $W_c \sim 3.5$ \cite{PhysRevB.82.174411}, which is later refined to $W_c \sim 3.7$ by finite size scaling \cite{PhysRevB.91.081103}.

In our study, we focus exclusively on the 1D lattice with $N = 16$ sites under periodic boundary conditions to eliminate boundary-induced localization effects. The many-body Hamiltonian is constructed on the basis of occupation number, where particle-number conserving terms such as $n_i$ result in a block-diagonal structure in the whole $2^{16}$ dimension space. We restrict our analysis to the half-filling sector with a total particle number of $N_f = 8$ fermions, for which the dimension of the Hilbert space is reduced to $D = \binom{16}{8} = 12870$.

\section{\label{Machine Learning Frameworks}Machine Learning Frameworks}

\subsection{\label{kan}Kolmogorov-Arnold Networks (KAN)}


The Kolmogorov-Arnold Networks (KAN) \cite{liu2025kan}, named after the mathematicians Andrey Kolmogorov and Vladimir Arnold, are inspired by the Kolmogorov-Arnold representation theorem \cite{kolmogorov1957representation,arnold1957functions}, which expresses multivariate functions as compositions of univariate functions and addition, offering a formulation more aligned with symbolic representations than the universal approximation theorem underlying traditional Multi-Layer Perceptrons (MLPs). Although this theorem has historically been considered of limited practical relevance for learning due to potential non-smoothness \cite{10.1162/neco.1989.1.4.465}, and prior studies remained largely restricted to shallow constructions \cite{6796921,SPRECHER200257}, Ref. \cite{liu2025kan} revisits this framework and introduces KAN as a fundamental redesign of MLPs. Specifically, KAN places learnable activation functions on edges ("weights") rather than nodes ("neurons") and generalizes the representation to deep and wide architectures. Each learnable weight parameter in an MLP is replaced by a learnable one-dimensional function (parameterized as a spline) in a KAN, while each node simply sums incoming signals without applying additional nonlinearities, leading to improved interpretability and favorable scaling behavior. Building upon this architecture, subsequent works have demonstrated its effectiveness in scientific machine learning; for example, the Kolmogorov–Arnold-Informed Neural Network (KINN) \cite{WANG2025117518} integrates KAN with physics-informed learning frameworks and achieves improved accuracy and efficiency in solving both forward and inverse partial differential equation problems.
\begin{figure}[tp]
\centering
\includegraphics[width=1.0\linewidth]{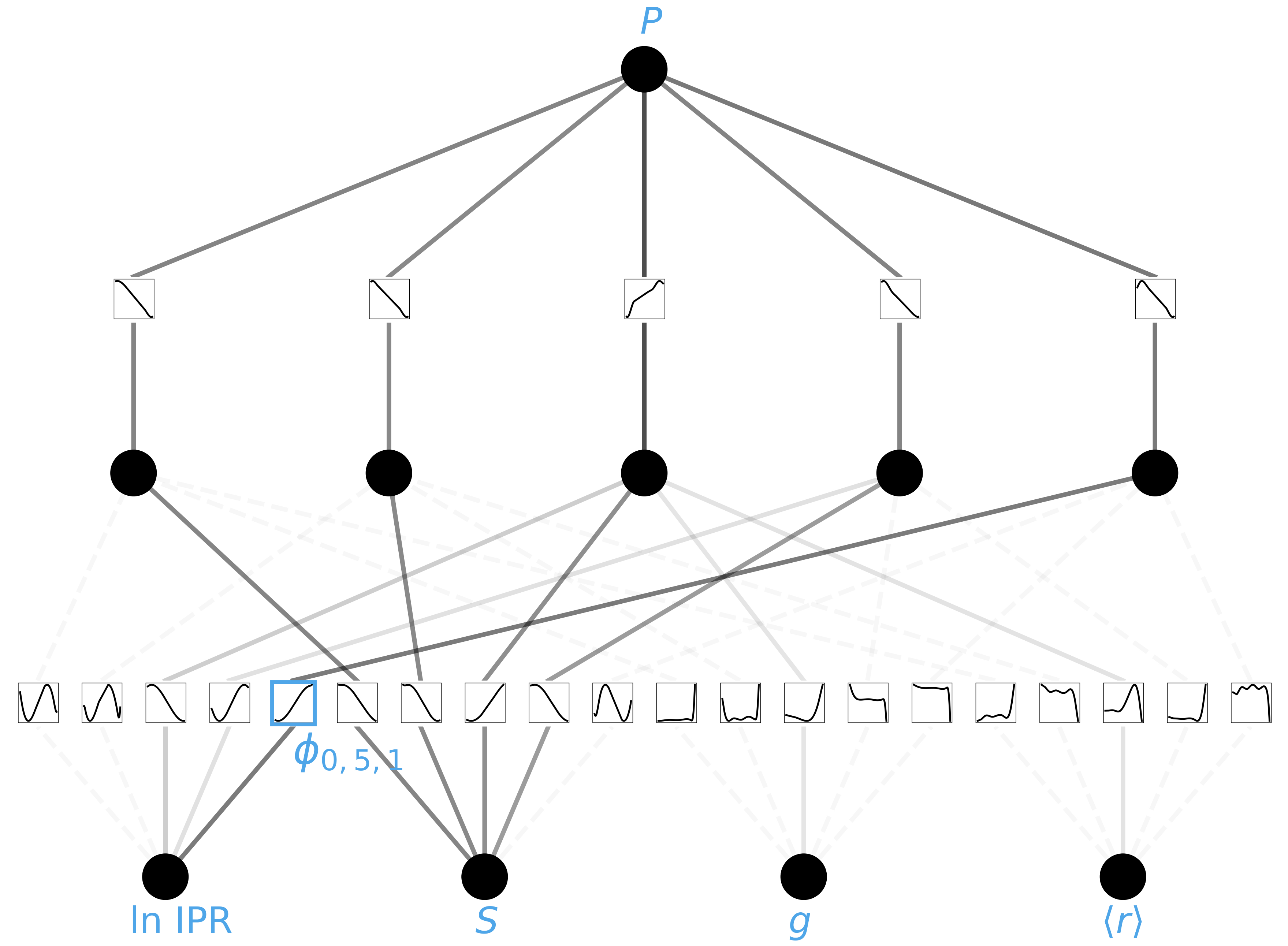}
\caption{Architecture of the KAN model with shape $[4,5,1]$. 
Learnable activation functions $\phi_{l,q,p}$, parameterized as B-spline curves, 
are associated with the edges rather than the nodes. The color intensity of each connection indicates its relative importance. Edges with importance scores below a threshold $3\times10^{-2}$ are shown as dashed lines. The complete set of importance score $Score$ is provided in Appendix \ref{score_table}.}
\label{fig:kan_arch}
\end{figure}

We illustrate in Fig.~\ref{fig:kan_arch} the architecture of our KAN model with shape $[4,5,1]$, i.e., 4 neurons in the input layer $l=0$, 5 nodes in the hidden layer $l=1$, and 1 neuron in the output layer $l=2$. The neuron $p$ in the $l$-th layer is labeled as $(l,p)$ and the learnable activation function $\phi_{l,q,p}$ is associated with the edge connecting node $(l,p)$ to node $(l+1,q)$. In this work, the architecture $[4,5,1]$ is specifically designed to process 4 physical features extracted from many-body wave functions, namely, the inverse participation ratio $\mathrm{IPR}$, the Shannon entropy $S$, many-body hybridization parameter $g$, and the mean level spacing ratio $\langle r \rangle $. The definition of these physical quantities is given in detail in Appendix \ref{app:features}. The network consists of two KAN layers:

\begin{itemize}
    \item \textbf{Layer $\Phi_0$}: Maps input features to hidden representations through a matrix $5 \times 4$ of learnable activation functions:
    \begin{equation}
    \Phi_0 = \begin{pmatrix}
    \phi_{0,1,1} & \cdots & \phi_{0,1,4} \\
    \vdots & \ddots & \vdots \\
    \phi_{0,5,1} & \cdots & \phi_{0,5,4}
    \end{pmatrix},
    \end{equation}
    
    \item \textbf{Layer $\Phi_1$}: Maps hidden representations to output through a matrix $1 \times 5$ of learnable activation functions:
    \begin{equation}
    \Phi_1 = \begin{pmatrix}
    \phi_{1,1,1} & \cdots & \phi_{1,1,5}
    \end{pmatrix}.
    \end{equation}
\end{itemize}
The forward pass of our neural network can be viewed as a mathematical instantiation of the Kolmogorov-Arnold representation, realized through successive layers of summation and activation function application
\begin{equation}
x_{1,q} = \sum_{p=1}^{4} \phi_{0,q,p}(x_{0,p}), \quad \text{for } q=1,\ldots,5, 
\end{equation}
and
\begin{equation}
x_{2,1} = \sum_{p=1}^{5} \phi_{1,1,p}(x_{1,p}),
\end{equation}
where $x_{l,p}$ is the activation value of the $(l,p)$-neuron. This "function-as-weight" paradigm allows KAN to capture complex nonlinear relationships directly through the composition $z(\mathbf{x})=\text{KAN}(\mathbf{x}) = (\Phi_1 \circ \Phi_0)(\mathbf{x})$, avoiding the need for deep stacking of nonlinear transformations as required in traditional MLP. All operations remain differentiable, enabling efficient training via backpropagation.

\begin{figure}[tbp]
\centering
\includegraphics[width=0.5\textwidth]{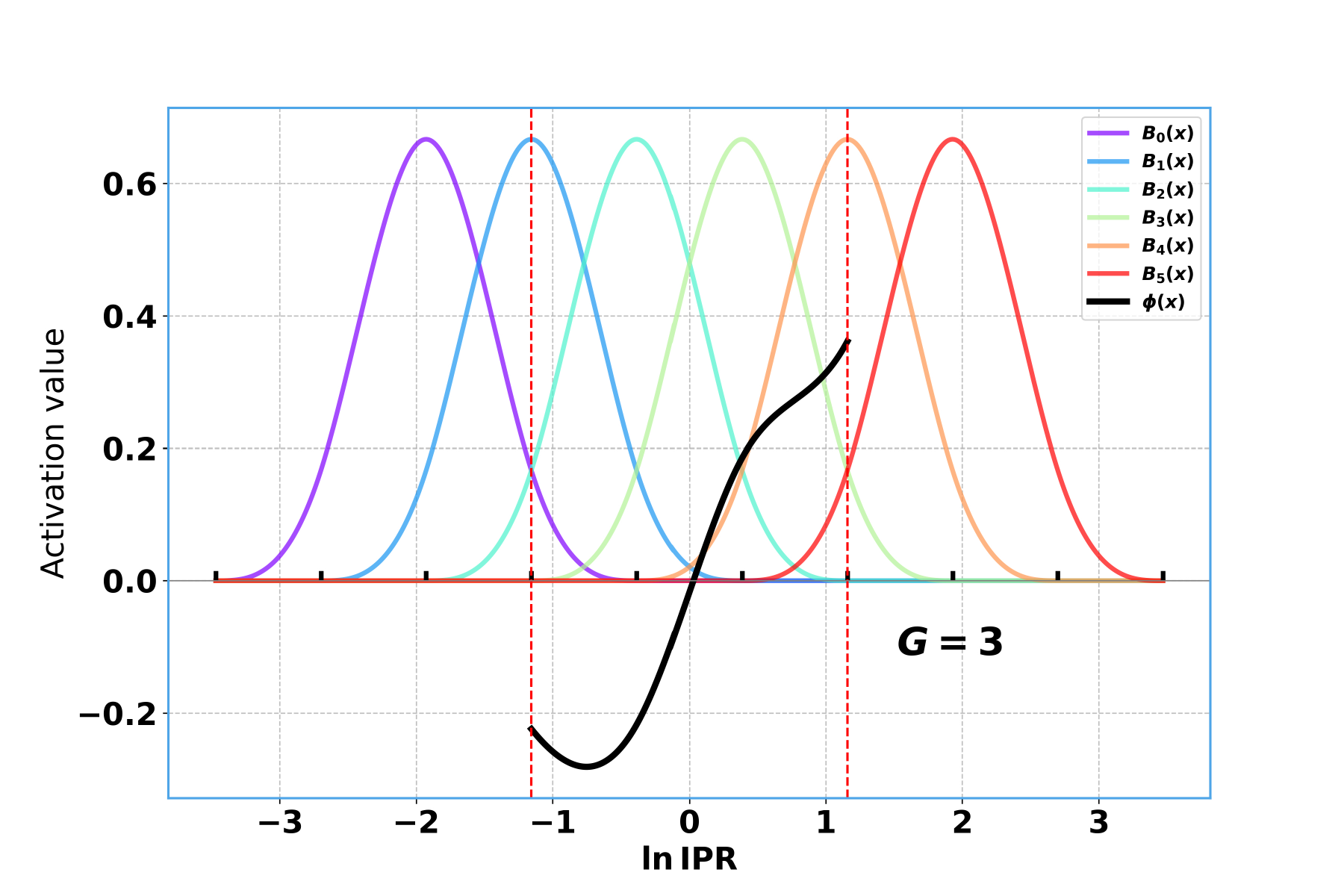}
    \caption{Illustration of the B-spline construction for the target function $\phi_{0,5,1}$ (solid black curve), which is computed via Eq.~(\ref{eq_func}). The six colored dashed curves correspond to the six B-spline basis functions whose weighted sum forms $\phi_{0,5,1}$. The horizontal axis represents the values of the feature $\ln{\mathrm{IPR}}$, while the vertical axis denotes the function values. The overall spline is defined on a grid of size $G=3$ with order $k=3$.}
    \label{fig:spline_1_3}
\end{figure}

We present a detailed analysis of the learned representation of the activation function $\phi_{0,5,1}$ between the first feature (the inverse participation ratio $\ln{\mathrm{IPR}}$) and the fifth neuron in the hidden layer in Fig.~\ref{fig:spline_1_3}. Each activation function, collectively denoted $\phi(x)$, is expressed as the sum of the basis function $b(x)$ (similar to residual connections) and a spline function, i. e.
\begin{equation}\label{eq_func}
\phi(x) = w_b b(x) + w_s \text{spline}(x),
\end{equation}
where
\begin{equation}\label{silu}
b(x) = \text{silu}(x) = \frac{x}{1 + e^{-x}}.
\end{equation}
The spline$(x)$ here is parametrized as a linear combination of B-spline basis functions $B_{i,k}$ of order $k$ defined on a grid of size $G$,
\begin{equation}
\text{spline}(x) = \sum_{i=0}^{G+k-1} c_i B_{i,k}(x). 
\end{equation}
$B_{i,k}$ is related to the lower order basis functions by the recursion relation
\begin{equation}
B_{i,k}(x)= \frac{x - t_i}{t_{i+k} - t_i}\, B_{i,k-1}(x) + \frac{t_{i+k+1} - x}{t_{i+k+1} - t_{i+1}} \, B_{i+1,k-1}(x),
\end{equation}
with the zeroth order functions given by
\begin{equation}
B_{i,0}(x) =
\begin{cases}
1, & t_{i} \le x < t_{i+1}, \\
0, & \text{otherwise},
\end{cases}
\end{equation}
for $i=0,\ldots,G+2k-1$. Here $w_b$, $w_s$, and $c_i$ are trainable parameters, and $\{t_0,\dots,t_{G+2k}\}$ is a non-decreasing knot sequence that defines the piecewise polynomial intervals. In this paper, we take $k=3$ and $G=3$. The six basis functions $B_i(x)$ and nine intervals separated by ten short bars are depicted in Fig.~\ref{fig:spline_1_3}. 

\begin{figure}[tp]
\centering
\includegraphics[width=1.0\linewidth]{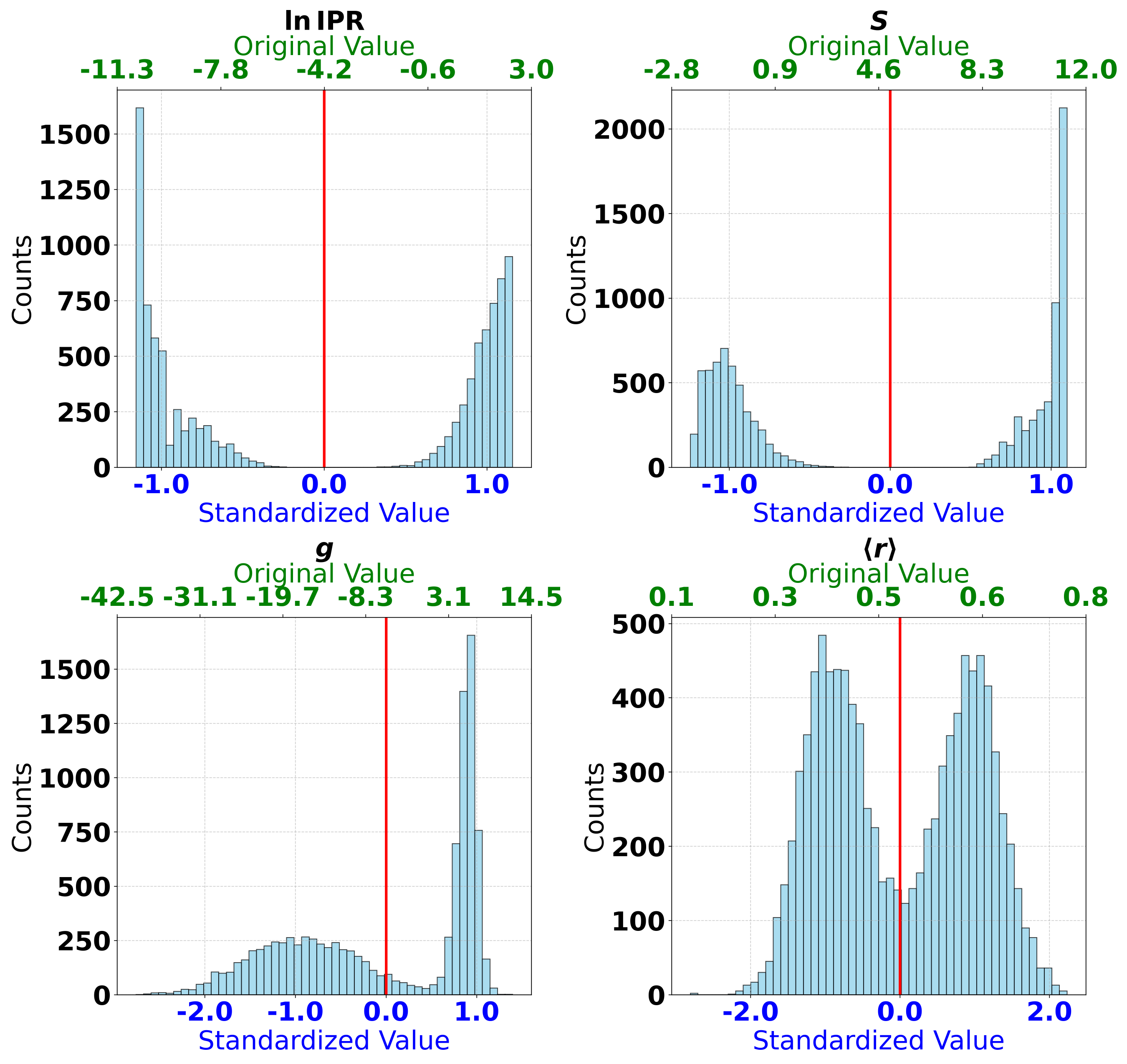}
\caption{Distribution of the four selected physical features $\ln \mathrm{IPR}$, $S$, $g$ and $\langle r\rangle$, before and after the standardization. Standardization centers each distribution at zero and scales it to unit variance, preserving its intrinsic shape.}
\label{fig:input_data}
\end{figure}

\subsection{\label{input_data}Input data}

We use four physically motivated quantities extracted from many-body wave functions as input features: $\ln \mathrm{IPR}$, $S$, $g$, and $\langle r\rangle$. Precise definitions and numerical procedures for their evaluation are provided in Appendix~\ref{app:features}.
This feature-based representation is economical in two respects. First, each wave function is mapped to a low-dimensional set of descriptors, which enables efficient data handling without storing the full probability distribution. Second, it significantly reduces the computational cost by avoiding more demanding observables, such as entanglement spectra. Importantly, the approach remains data-driven: no assumptions are imposed on the microscopic structure of the phases, and the relevant discriminative patterns are learned directly from the data.

For training neural networks to differentiate the ergodic and MBL wave functions, we select representative disorder strengths deep within each phase: $(W_\mathrm{ergodic}, W_\mathrm{MBL}) = (0.2, 12)$ for the 1D lattice. Choosing these values does not require knowledge of the critical value of disorder $ W_c $, because one can always verify that conventional observables follow the expected ergodic/MBL behaviors at these values. For the phase diagrams, we define a suitable grid of $W$ values in the intermediate region: $W \in [0.2, 4.6]$. At each selected $W$ value, we implement 50 random disorder realizations and perform exact diagonalization of the Hamiltonian. The energy spectrum is normalized as the energy density $ \epsilon = (E - E_\mathrm{min})/(E_\mathrm{max} - E_\mathrm{min})$ with  $E_\mathrm{min}(E_\mathrm{max})$ the smallest (largest) eigenvalue of the spectrum. The eigenstates of each disorder realization are binned into 20 equal energy intervals between $\epsilon = 0$ and $1$. In each bin, we discard all but the 50 eigenstates with energy densities closest to the center of the bin, greatly reducing data storage and computational demands during subsequent analysis. However, we do keep all the eigenvalues for computing the energy-level statistics later. We observe that using such a small sample of eigenstates does not significantly affect the disorder-averaged values of IPR and machine predictions. Note that, due to the low density of states near $\epsilon = 0$ and $1$, bins in these regions may contain fewer than 50 eigenstates per disorder realization.

Each time we train the model, we prepare a labeled dataset by randomly selecting 10,000 ergodic and 10,000 MBL wave functions with energy densities $0.15 < \varepsilon < 0.85$ \cite{PhysRevB.82.174411}, computed at disorder strengths $W_{\text{ergodic}}$ and $W_{\text{MBL}}$ as discussed above. From each wave function, we extract the four selected features and pair the resulting feature vector with a label of 1 (ergodic) or 0 (MBL). Compared to using full wave functions, this significantly simplifies the classification task and makes it more interpretable.

The histogram distribution and scaling of the four features are presented in Fig.~\ref{fig:input_data}, where a process of standardization is applied to center and normalize the distributions without altering their intrinsic shape, as evident from the unchanged histogram profiles. Within each panel, histograms show the frequency distribution of the feature values across the entire dataset (20,000 wave functions). The same distribution is displayed with dual horizontal axes: the top axis (green ticks) indicates the original (raw) feature range, while the bottom axis (blue ticks) corresponds to the standardized values. The standardized values are obtained via $x_s = (x - \mu)/\sigma$, where $\mu$ is the mean value and $\sigma$ the standard deviation. The visualization of both axes highlights the original value ranges and demonstrates how standardization renders the features dimensionless and comparable in scale — a common preprocessing step that aids model training and interpretability.
In particular, the distribution of the $g$ factor exhibits a clear bimodal structure. From the histogram of $g$, one can observe that samples in the ergodic phase ($W=0.2$) and in the MBL phase ($W=12$) form two distinct peaks. The peak associated with the MBL phase is broader and lower, while peak for the ergodic phase is sharper and higher. A quantitative evaluation shows that for $L=16$ and $W=12$ the standard deviation of the distribution is approximately $\sigma \approx 7$, which is consistent with the results reported in Ref.~\cite{PhysRevX.5.041047}.

To classify the two phases, we employ two complementary approaches: KAN, which provide a flexible and interpretable framework for modeling the nonlinear relationships between tabular physical features, and CNN, which process raw spatial data by extracting hierarchical features through convolutional layers, serving as a comparative benchmark. While KAN operate directly on low-dimensional feature vectors, CNN process the raw, high-dimensional wave functions instead. In our context, the input to the KAN model is a 4-component feature vector $\mathbf{x}=(\ln \mathrm{IPR}, S, g, \langle r\rangle)$, while the input to CNN is a probability distribution of dimension $D$. The KAN model is trained to distinguish between the ergodic and MBL phases solely on the basis of these features, without going into the details of the wave function.

\subsection{\label{training}Model training}

The goal is to train a KAN to classify eigenstates into ergodic or MBL phases using the four features. 
Each sample is a standardized feature vector $1\times 4$ with binary label $y$, where $y=1(0)$ denotes ergodic(MBL) phase. Data are split into a training set ($50\%$) and a test set ($50\%$). The network is trained using mini-batch gradient descent with a batch size of $M$. Each connection in the $\mathrm{KAN}[4,5,1]$ architecture is represented by a learnable B-spline function with the initialized trainable parameters close to the identity. The probability of the state being in ergodic phase is then obtained by applying a sigmoid function $\sigma(\cdot)$ on the network output $z$ as $P(\mathbf{x})=\sigma(z)$.

\textbf{Loss function:}
For binary ergodic/MBL classification, the network produces an output $z_i$ for each input sample. We adopt \texttt{BCEWithLogitsLoss} in PyTorch \cite{NEURIPS2019_bdbca288}, which combines the sigmoid function and binary cross-entropy loss in a numerically stable form
\begin{equation}
L(\mathbf{x}) = -\frac{1}{M} \sum_{i=1}^M \left[ y_i \log \sigma(z_i) + (1-y_i) \log \left( 1 - \sigma(z_i) \right) \right],
\label{eq_loss}
\end{equation}
where $y_i \in \{0,1\}$ is the true ergodic/MBL label. This choice is crucial for our task for two reasons. First, the combined formulation ensures numerical stability when the network output $z_i$ takes large positive or negative values, which frequently occurs in high-accuracy regimes. Second, the probabilistic interpretation of $\sigma(z)$ as $P$ allows the trained model to provide a physically meaningful confidence level for phase classification, allowing for direct visualization of the ergodic–MBL boundary in the phase diagram, i.e., the many-body mobility edge in the spectrum.

\textbf{Gradient descent:}
The network is optimized with the Adam algorithm \cite{kingma2015adam} in mini-batches of $M$ samples. The gradient is back propagated through all spline coefficients $w_s$ and bias terms $w_b$. Each update step corresponds to one mini-batch, and training is performed for $2000$ update steps, which amounts to approximately $12$ epochs over the training dataset.

\begin{figure}[tp]
    \centering
    \includegraphics[width=0.45\textwidth]{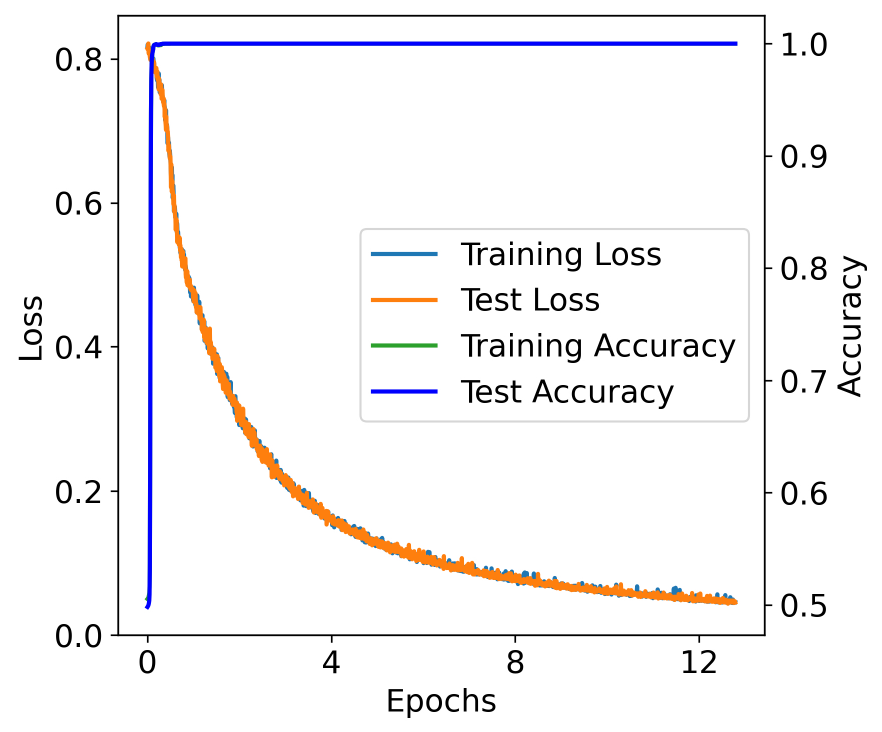}
    \caption{Training and test loss functions and accuracies of the KAN classifier.}

    \label{fig:kan_training}
\end{figure}

\textbf{Training history:}
During the training process, both the training and test losses exhibit a rapid initial decay followed by a saturation plateau. The negligible discrepancy between the two curves indicates robust generalization capability. The model achieves high accuracy throughout training, with test accuracy consistently exceeding $>99.95\%$. The evolution of the training loss, accuracy, and their corresponding test metrics is illustrated in Fig.~\ref{fig:kan_training}.

\subsection{\label{diagram}Phase Diagram Prediction}

To obtain the energy-resolved phase diagram of the one-dimensional interacting fermionic chain, we trained two supervised classifiers. The first is a Kolmogorov–Arnold Network (KAN) constructed from four physical observables, $\ln \mathrm{IPR}$, $S$, $g$, and $\langle r\rangle$. The second is a convolutional neural network (CNN) operating directly on the full probability distributions of many-body eigenstates; details are given in Appendix~\ref{cnn_appendix}. Both models achieve an average validation accuracy above $99.9\%$ on test states, indicating that they reliably distinguish states deep within the ergodic and MBL regimes.

Fig.~\ref{fig:phase-comparison} compares the phase diagrams obtained from the conventional numerical benchmark and the two machine-learning approaches. Panel~(a) shows the result based on $-\ln \mathrm{IPR}$, while panels~(b) and (c) display the predictions of the KAN and CNN classifiers, respectively. The color scale represents the averaged network output $\langle P\rangle$, where the average is taken over eigenstates within each energy window and over disorder realizations. Values $\langle P\rangle\approx1$ correspond to the ergodic phase, and values $\langle P\rangle\approx0$ indicate MBL characteristics.

Both machine-learning models reproduce the global structure of the phase diagram, including the energy-dependent bending of the many-body mobility edge and the separation between ergodic and MBL regions. The crossover line around $W_c\sim2.8$ \cite{PhysRevB.109.075124,PhysRevB.95.245134} is consistently captured. Within numerical resolution, the KAN and CNN predictions are nearly indistinguishable over most of the spectrum, indicating that both feature-based and raw-data-based learning can reconstruct the phase structure with high accuracy.

\begin{figure}[tp]
    \centering
    \begin{subfigure}[t]{0.23\textwidth}
        \centering
        \begin{overpic}[width=\textwidth]{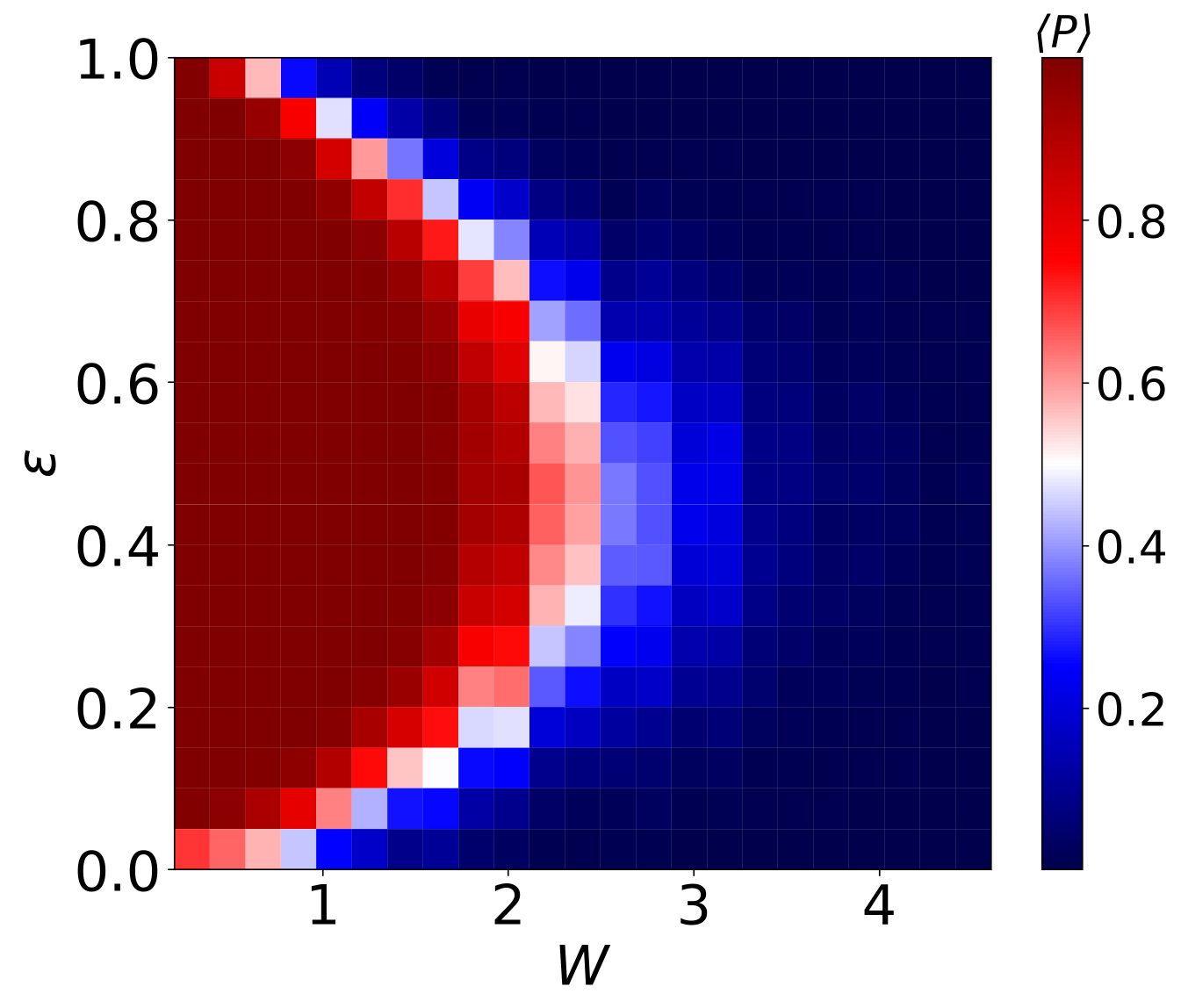}
          \put(70,70){\fontsize{12}{14}\selectfont\color{white}(a)}
        \end{overpic}
        \label{fig:sub1}
    \end{subfigure}
    \begin{subfigure}[t]{0.23\textwidth}
        \centering
        \begin{overpic}[width=\textwidth]{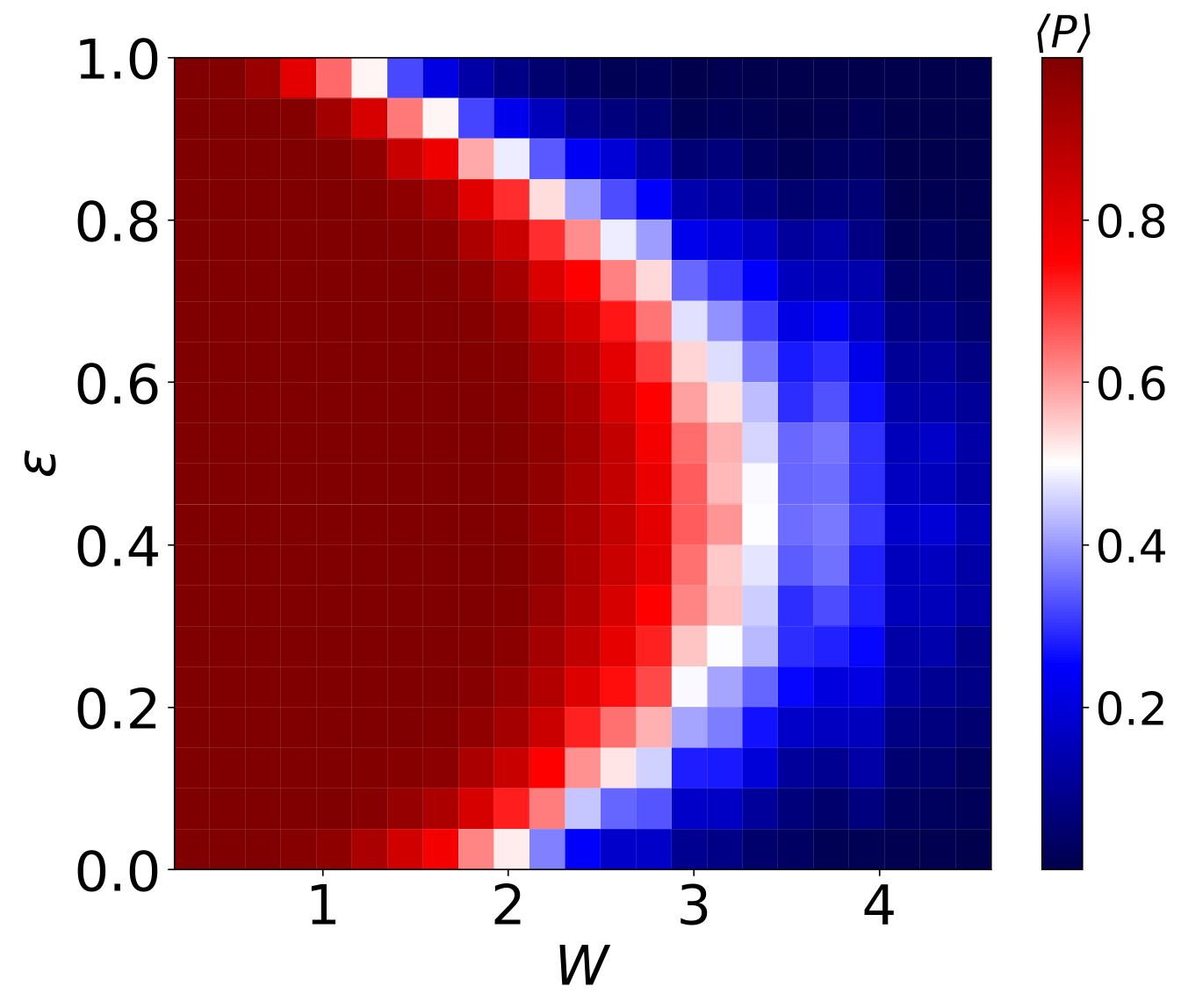}
         \put(70,70){\fontsize{12}{14}\selectfont\color{white}(b)}
        \end{overpic}
        \label{fig:sub2}
    \end{subfigure}
    \medskip
    \begin{subfigure}[t]{0.23\textwidth}
        \centering
        \begin{overpic}[width=\textwidth]{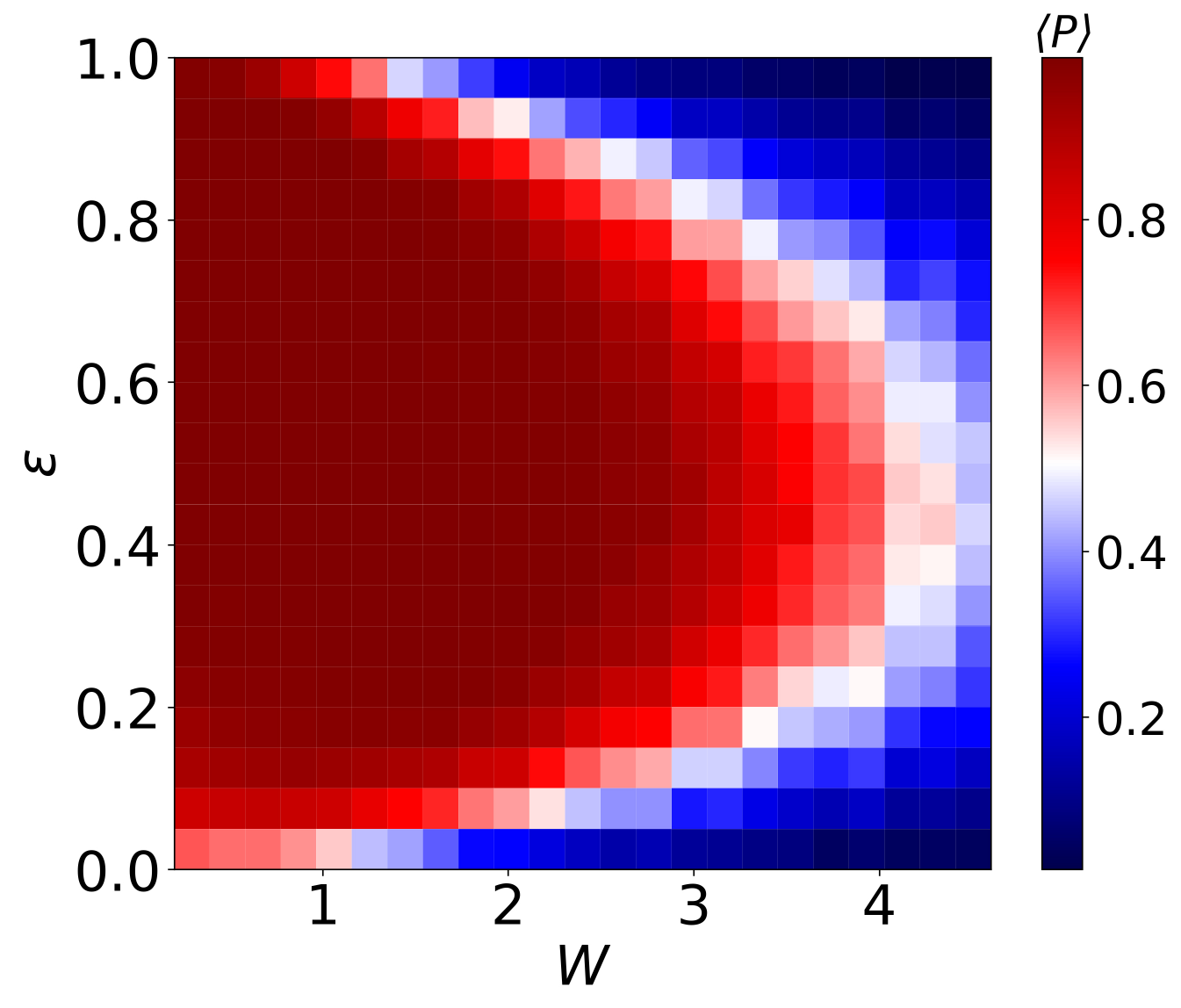}
         \put(70,70){\fontsize{12}{14}\selectfont\color{white}(c)}
        \end{overpic}
        \label{fig:sub3}
    \end{subfigure}
    \begin{subfigure}[t]{0.23\textwidth}
        \centering
        \begin{overpic}[width=\textwidth]{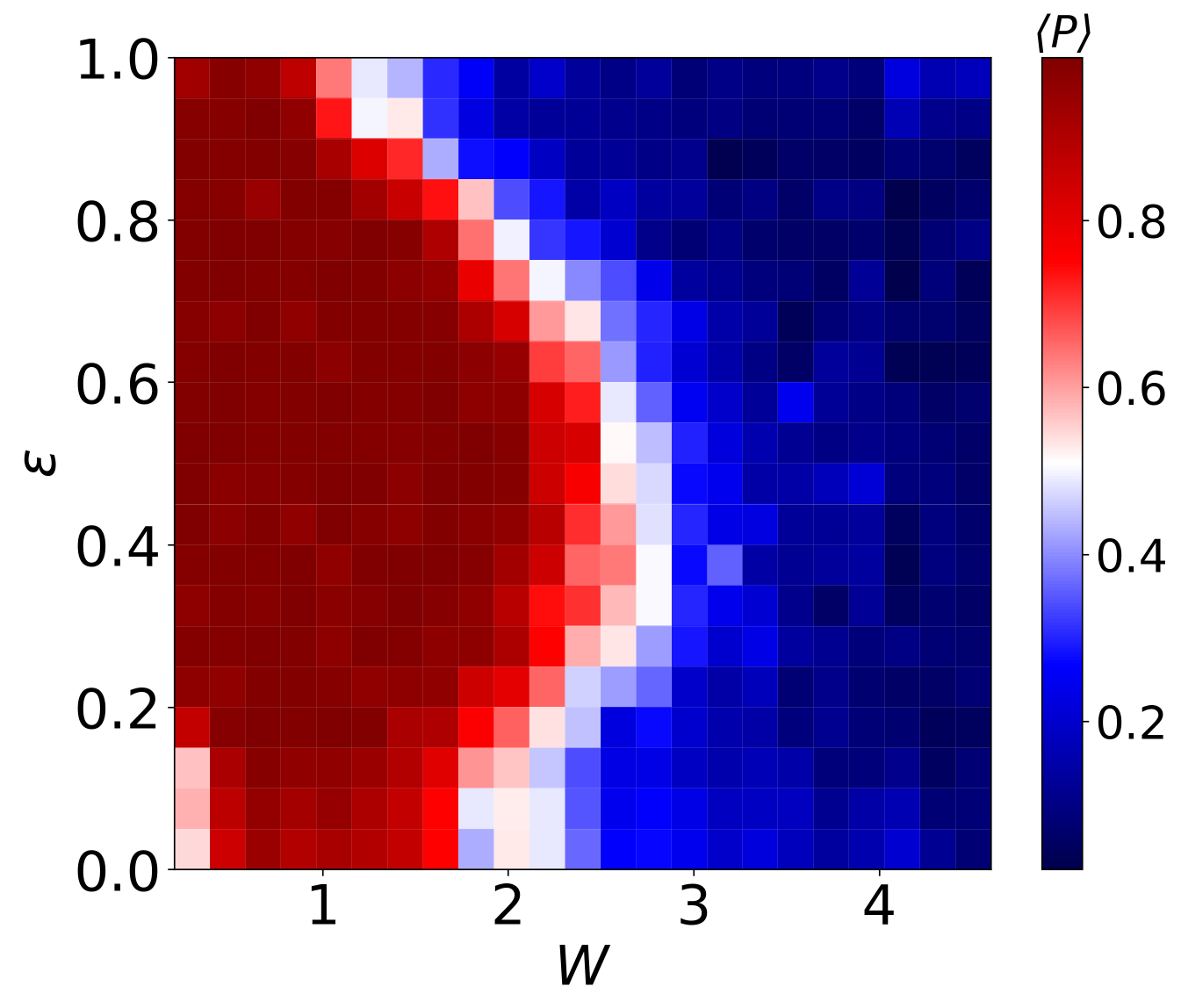}
         \put(70,70){\fontsize{12}{14}\selectfont\color{white}(d)}
        \end{overpic}
        \label{fig:sub4}
    \end{subfigure}

    \caption{Phase diagram predicted by the KAN classifier trained independently on (a) $\ln \mathrm{IPR}$, (b) $S$, (c) $g$, and (d) $\langle r\rangle$, respectively.}
    \label{fig:all_four}
\end{figure}

Fig.~\ref{fig:all_four} further compares the phase diagrams predicted by KAN models trained independently on each observable. The inferred transition points are not consistent across features: models based on $S$ and $g$ tend to predict a relatively larger critical disorder strength, whereas those based on $\ln \mathrm{IPR}$ and $\langle r\rangle$ systematically yield smaller estimates. This discrepancy indicates that different observables probe distinct aspects of the transition, motivating a combined treatment in which all four features are incorporated to obtain a more balanced and reliable phase diagram. To quantify their respective contributions, we introduce edge-importance scores for all connections in the trained KAN classifier with architecture $[4,5,1]$ (see Appendix \ref{score_table}). The detailed statistics summarized in Table~\ref{tab:edge_scores_all} show that $\ln \mathrm{IPR}$ and $S$ have a larger number of edges with importance scores exceeding the threshold value $3\times10^{-2}$, indicating that these two observables play a more prominent role in shaping the decision function.


\section{\label{conclusion}CONCLUSIONS}

In this work, we investigated the ergodic-MBL transition in a one-dimensional interacting disordered fermionic system by combining exact diagonalization with supervised machine learning. Two complementary approaches were implemented: a feature-based Kolmogorov–Arnold Network (KAN) built upon four physically motivated observables and a convolutional neural network (CNN) trained directly on high-dimensional wave-function probability densities.

Both approaches achieve high classification accuracy deep in the ergodic and MBL phases and reproduce the energy-resolved phase diagram with quantitative consistency. The comparison demonstrates that incorporating multiple physical observables leads to more stable and sharper phase-boundary identification than relying on any single feature alone. The joint use of $\ln \mathrm{IPR}$, $S$, $g$, and $\langle r\rangle$ enables robust classification across the entire spectrum, highlighting the advantage of combining complementary physical information.

At the same time, the KAN framework exhibits a clear computational advantage. Owing to its low-dimensional structured input and explicit functional decomposition, the KAN model trains approximately an order of magnitude faster than the CNN while achieving comparable predictive performance. Moreover, the explicit edge functions and importance scores provide direct interpretability of how different physical quantities contribute to the learned decision function.

Beyond the present system, the KAN architecture offers a flexible and extensible framework. Additional physically motivated observables can be incorporated straightforwardly as new input nodes without altering the overall structure, allowing systematic integration of richer physical information. This scalability, together with its computational efficiency and interpretability, makes the feature-based KAN approach a promising tool for phase identification and analysis in a broad class of quantum many-body systems.

\begin{acknowledgments}
This work is supported by the Natural Science Foundation of China (Grants No. 12474492, No. 12461160324, No. 12404285, No. 12404322), the Science Challenge Project (Grant No. TZ2025017), the Zhejiang Provincial Natural Science Foundation of China (Grant No. LQ24A040004, LQN25A040003), and the Science Foundation of
Zhejiang Sci-Tech University (Grant No. 23062152-Y, No. 23062182-Y).
\end{acknowledgments}

\appendix
\section{\label{app:JW}Jordan--Wigner Transformation}

In this appendix, we present a complete derivation of the mapping from the interacting
spinless fermion Hamiltonian in Eq.~(\ref{eq1}) to the spin-$\frac{1}{2}$ Heisenberg model
in Eq.~(\ref{eq2}) using the Jordan--Wigner transformation.

In one dimension, fermionic creation and annihilation operators can be mapped exactly
onto spin-$\frac{1}{2}$ operators as
\begin{equation}
\label{eq:JW}
\begin{aligned}
c_i &= \exp\!\left(i\pi \sum_{k<i} n_k\right) S_i^- , \\
c_i^\dagger &= S_i^+ \exp\!\left(-i\pi \sum_{k<i} n_k\right)  , \\
n_i &= c_i^\dagger c_i = S_i^z + \frac{1}{2},
\end{aligned}
\end{equation}
The nonlocal string operator ensures the correct fermionic anticommutation relations. The inverse Jordan--Wigner transformation expresses the spin operators in terms of fermionic operators,
\begin{equation}
\label{eq:JW_inverse}
\begin{aligned}
S_i^+ &= c_i^\dagger \exp\!\left(i\pi \sum_{k<i} n_k\right), \\
S_i^- &= \exp\!\left(-i\pi \sum_{k<i} n_k\right) c_i , \\
S_i^z &= n_i - \frac{1}{2}.
\end{aligned}
\end{equation}
Together, Eqs.~(\ref{eq:JW}) and (\ref{eq:JW_inverse}) establish an exact mapping between the interacting spinless fermion model and the spin-$\frac{1}{2}$ chain in one dimension.

For nearest-neighbor sites $j=i+1$, the Jordan--Wigner strings cancel exactly, yielding
\begin{equation}
\label{eq:JW_hopping}
c_i^\dagger c_{i+1} + c_{i+1}^\dagger c_i
=
S_i^+ S_{i+1}^- + S_i^- S_{i+1}^+ .
\end{equation}
Using the identity
\begin{equation}
S_i^+ S_j^- + S_i^- S_j^+
=
2 \left( S_i^x S_j^x + S_i^y S_j^y \right),
\end{equation}
the hopping term in Eq.~(\ref{eq1}) becomes
\begin{equation}
- t \sum_{\langle i,j\rangle}
\left( c_i^\dagger c_j + c_j^\dagger c_i \right)
=
-2t \sum_{\langle i,j\rangle}
\left( S_i^x S_j^x + S_i^y S_j^y \right).
\end{equation}
Using the relation $n_i - \tfrac{1}{2} = S_i^z$, the nearest-neighbor interaction term is mapped into
\begin{equation}
V \sum_{\langle i,j\rangle}
\left(n_i - \frac{1}{2}\right)
\left(n_j - \frac{1}{2}\right)
=
V \sum_{\langle i,j\rangle}
S_i^z S_j^z .
\end{equation}
The term with on-site disorder is transformed into a random magnetic field along the $z$-axis,
\begin{equation}
\sum_i u_i \left(n_i - \frac{1}{2}\right)
=
\sum_i u_i S_i^z .
\end{equation}

Collecting all contributions, the Hamiltonian becomes
\begin{equation}
H =
\sum_{\langle i,j\rangle}
\left[
-2t \left( S_i^x S_j^x + S_i^y S_j^y \right)
+ V S_i^z S_j^z
\right]
+ \sum_i u_i S_i^z .
\end{equation}
The negative sign in front of the transverse ($XY$) coupling originates
from the fermionic hopping term and indicates a ferromagnetic interaction
in the $x$--$y$ plane. In one dimension, this sign can be removed by a
staggered unitary transformation,
$S_i^{x,y} \rightarrow (-1)^i S_i^{x,y}$ and $S_i^z \rightarrow S_i^z$,
which leaves the commutation relations and the $S^z$ sector invariant,
while flipping the sign of the nearest-neighbor $XY$ term.
After this transformation, the Hamiltonian takes the standard antiferromagnetic form.

At the isotropic point $t=\tfrac{1}{2}$ and $V=1$, this reduces to
\begin{equation}
H =
\sum_{\langle i,j\rangle}
\mathbf{S}_i \cdot \mathbf{S}_j
+ \sum_i u_i S_i^z ,
\end{equation}
which is identical to Eq.~(\ref{eq2}) in the main text. At half filling, the conservation of the fermionic particle number corresponds to zero total magnetization $\sum_i S_i^z = 0$, which defines the sector studied in this work.

\section{Feature definitions and numerical implementation}
\label{app:features}

In this Appendix, we provide the explicit definitions and numerical procedures used to compute the four input features employed in the KAN classification.

\textbf{Wave-function representation:}
All features are computed from exact many-body eigenstates obtained by full diagonalization of the Hamiltonian. 
For system size $L=16$ at half filling, the Hilbert space dimension is $D=\binom{16}{8}$. 
Each normalized eigenstate is expanded in the occupation-number basis as
\begin{equation}
\ket{\psi_n} = \sum_{\alpha=1}^{D} \psi_{n,\alpha} \ket{\alpha},
\end{equation}
with associated probability distribution
\begin{equation}
p_{n,\alpha} = |\psi_{n,\alpha}|^2,
\qquad
\sum_{\alpha=1}^{D} p_{n,\alpha} = 1 .
\end{equation}

\textbf{Inverse participation ratio:}
The inverse participation ratio (IPR), which characterizes the localization of the wave function in Hilbert space, is defined as
\begin{equation}
\mathrm{IPR}_n = \sum_{\alpha=1}^{D} p_{n,\alpha}^2 .
\end{equation}
For an ergodic state uniformly spread over the Hilbert space, $\mathrm{IPR}_n \sim 1/D$, whereas for a localized state it remains finite. 
To reduce the wide dynamical range and improve numerical stability, we use $\ln(\mathrm{IPR}_n)$ as an input feature.

\textbf{Shannon entropy:}
The Shannon entropy of an eigenstate is defined as
\begin{equation}
S_n = -\sum_{\alpha=1}^{D} p_{n,\alpha} \ln p_{n,\alpha} .
\end{equation}
This quantity measures the information-theoretic spread of the wave function over the many-body basis. 
In the ergodic phase, $S_n$ approaches a value close to $\ln D$, whereas in the MBL phase it is substantially reduced, reflecting the localized structure of the eigenstates.

\textbf{Many-body hybridization parameter $g$:}
To quantify the hybridization between adjacent many-body eigenstates in the occupation-number basis introduced above, we construct a Thouless-type parameter using matrix elements of local density operators.

Let $\{E_n\}$ denote the ordered many-body spectrum with $E_{n+1} > E_n$, and define the adjacent level spacing
\begin{equation}
\delta_n = E_{n+1} - E_n .
\end{equation}
Since the local density operator $\hat n_i$ is diagonal in the occupation-number basis $\{ \ket{\alpha} \}$, its off-diagonal matrix element between neighboring eigenstates reads
\begin{equation}
\langle \psi_n | \hat n_i | \psi_{n+1} \rangle
=
\sum_{\alpha=1}^{D}
\psi_{n,\alpha}^*
\psi_{n+1,\alpha}
\, n_i(\alpha),
\end{equation}
where $n_i(\alpha)\in\{0,1\}$ denotes the occupation number of site $i$ in configuration $\alpha$.
For each site $i$, we define a site-resolved hybridization parameter
\begin{equation}
g_n^{(i)}
=
\ln
\left(
\frac{
\left|
\langle \psi_n | \hat n_i | \psi_{n+1} \rangle
\right|
}{
\delta_n
}
\right).
\end{equation}

To reduce site-dependent fluctuations, we average over a set of spatially separated lattice sites $\mathcal{S}$,
\begin{equation}
g_n
=
\frac{1}{N_s}
\sum_{i \in \mathcal{S}}
g_n^{(i)}.
\end{equation}
In the numerical implementation, we choose $\mathcal{S}=\{0, L/4, L/2, 3L/4\}$ for a chain of length $L$, such that $N_s=4$.

The parameter $g_n$ directly compares the magnitude of off-diagonal matrix elements to the corresponding level spacing. 
In the ergodic phase, local matrix elements are comparable to the level spacing, yielding finite or positive values of $g_n$. 
In contrast, in the MBL phase, local matrix elements are exponentially suppressed relative to the level spacing, resulting in negative $g_n$. 
Thus, the distribution of $g_n$ provides a sensitive probe of the ergodic-MBL transition.

\textbf{Mean level spacing ratio:}
Spectral correlations are incorporated through the level spacing ratio. 
For the ordered many-body spectrum $\{E_n\}$ introduced above, with level spacings $\delta_n$, we define
\begin{equation}
r_n =
\frac{\min(\delta_n, \delta_{n-1})}
{\max(\delta_n, \delta_{n-1})}.
\end{equation}
For each eigenstate, we compute a local mean spacing ratio by averaging $r_n$ within a finite energy window centered at that state. This procedure suppresses edge effects and reduces finite-size fluctuations.

In the ergodic phase, the local mean ratio approaches the Wigner--Dyson value ($0.53$) characteristic of chaotic spectra, whereas in the MBL phase it approaches the Poisson value $(0.386)$ expected for localized systems.

\section{\label{cnn_appendix}Convolutional Neural Network (CNN)}

The CNN employed in this work consists of the following components: a single one-dimensional convolutional layer with 16 kernels of length $l=10$, each followed by a rectified linear unit (ReLU) activation function defined as $\mathrm{ReLU}(x)=\max(0,x)$; a maximum pooling layer with pooling size 2; a fully connected layer comprising 60 neurons with ReLU activation; a dropout layer with dropout rate $d=0.2$; and a final output layer with a single sigmoid neuron that yields the probability $P$ of classification into ergodic phase.

CNN takes as input the full many-body wave functions, which constitute a Hilbert-space of dimension 12,870 and are represented as one-dimensional vectors of probability densities $|\psi_\alpha|^2$, where $\alpha$ labels the occupation-number basis states. These vectors are normalized and directly supplied to the convolutional layer without explicit feature extraction. Following Chen \cite{PhysRevB.109.075124}, this representation can be viewed as a one-dimensional grayscale image of size $1\times12{,}870$, in which each element corresponds to the probability associated with a specific basis state. While this encoding preserves the complete information of the wave functions, it requires processing high-dimensional data, which poses significant computational challenges.

Model training is performed using mini-batch gradient descent with a batch size of 50. The optimization proceeds over multiple epochs, with early stopping applied when the validation loss no longer decreases. To mitigate the effects of random weight initialization and stochastic data sampling, we adopt a 20-fold cross-validation scheme, training 20 independent CNN models and reporting statistically averaged results, consistent with the procedure described in Ref.~\cite{PhysRevB.109.075124}.

\section{\label{score_table}Edge importance scores}

This appendix lists the importance scores assigned to all edges of the trained
KAN classifier with architecture $[4,5,1]$.
The scores quantify the relative contribution of individual connections to the
model output, as discussed in the main text.
Edges with importance scores below $3\times10^{-2}$ are indicated as dashed lines
in Fig.~\ref{fig:kan_arch}.

In the network, each connection between neurons is not represented by a scalar weight, but by an independent, learnable, one-dimensional function $\phi_{l,q,p}$. To quantify the contribution of each edge to the final classification outcome, we employ a backward attribution metric implemented in the KAN framework.

For a given edge connecting neuron $p$ in layer $l$ to neuron $q$ in layer $l+1$, the importance score ($Score$) is defined as the magnitude of the gradient of the loss function with respect to the corresponding activation function, averaged in the data set
\begin{equation}
Score_{l,p\to q}
=
\mathbb{E}_{\mathbf{x}\sim\mathcal{D}}
\left[
\left|
\frac{\partial L(\mathbf{x})}{\partial \phi_{l,q,p}}
\right|
\right],
\end{equation}
where $\mathbf{x}$ denotes an input sample drawn from the data distribution $\mathcal{D}$, and $L(\mathbf{x})$ is the binary cross-entropy loss evaluated on that input, defined as in Eq.~(\ref{eq_loss}). The expectation value $\mathbb{E}_{\mathbf{x}\sim\mathcal{D}}[\cdot]$ indicates an average over all input data in the distribution; in practice, it is approximated by the arithmetic mean over all samples in the training or validation dataset. Taking the absolute value removes sign cancellations and ensures that the score captures the overall sensitivity of the loss to variations in the functional form of the edge activation. Consequently, the importance score quantifies the global influence of an edge by measuring how strongly perturbations of its associated activation function affect the network output across the dataset, rather than reflecting the effect of a single scalar parameter.

The resulting values of $Score$ exhibit a highly heterogeneous distribution that spans several orders of magnitude.
As summarized in Table~\ref{tab:edge_scores_all}, only a small subset of edges carry scores of order unity, corresponding to dominant information pathways that effectively act as order-parameter channels for the ergodic-MBL classification.
In particular, the connections $(1\!\to\!5)$ in Layer~0 and $(3\!\to\!1)$ in Layer~1 exhibit the highest scores, while most of the remaining edges have vanishingly small values.
This pronounced sparsity reflects both the physics-informed choice of input features and the expressive bias of the KAN architecture.
For visualization purposes, the raw scores listed in Table~\ref{tab:edge_scores_all} are mapped to edge transparencies using a monotonic function $\tanh$, and connections below a fixed threshold $3\times 10^{-2}$ are shown as dashed lines.

\begin{table}[htbp]
\centering
\caption{Importance scores of all edges in the KAN classifier.}
\label{tab:edge_scores_all}

\begin{tabular}{ccS[table-format=1.10]} 
\hline
Input & Output & \multicolumn{1}{c}{$Score$} \\
\hline
\multicolumn{3}{c}{Layer 0} \\
\hline
1 & 1 & \textcolor{gray}{0.0058150650} \\
1 & 2 & \textcolor{gray}{0.0007361337} \\
1 & 3 & 0.0643463954 \\
1 & 4 & 0.0391842275 \\
1 & 5 & 0.1913769245 \\
2 & 1 & 0.1740498691 \\
2 & 2 & 0.1674466878 \\
2 & 3 & 0.1575163453 \\
2 & 4 & 0.1363180279 \\
2 & 5 & \textcolor{gray}{0.0103436997} \\
3 & 1 & \textcolor{gray}{0.0006724185} \\
3 & 2 & \textcolor{gray}{0.0003671655} \\
3 & 3 & 0.0343550741 \\
3 & 4 & \textcolor{gray}{0.0003630245} \\
3 & 5 & \textcolor{gray}{0.0020129866} \\
4 & 1 & \textcolor{gray}{0.0016932463}\\
4 & 2 & \textcolor{gray}{0.0003827065} \\
4 & 3 & 0.0367036387\\
4 & 4 & \textcolor{gray}{0.0056185782} \\
4 & 5 & \textcolor{gray}{0.0000510799} \\
\midrule
\multicolumn{3}{c}{Layer 1} \\
\midrule
1 & 1 & 0.1745929569 \\
2 & 1 & 0.1676083803 \\
3 & 1 & 0.2890724242 \\
4 & 1 & 0.1754709482 \\
5 & 1 & 0.1941169202 \\
\hline
\end{tabular}
\end{table}

\bibliography{KAN-PRB}

\end{document}